\documentclass[aps,pra,twocolumn,superscriptaddress,showpacs]{revtex4}

\usepackage{amsmath}
\usepackage{amstext}
\usepackage{amssymb}
\usepackage{bbm}
\usepackage{latexsym}
\usepackage[dvips]{graphicx}

\newcommand{\ket}[1]{\left\vert#1\right\rangle}
\newcommand{\bra}[1]{\left\langle#1\right\vert}
\newcommand{\Ham}{\mathcal H}

\begin{document}

\author{D. Rossini}
\affiliation{International School for Advanced Studies (SISSA),
  Via Beirut 2-4, I-34014 Trieste, Italy}

\author{P. Facchi}
\affiliation{Dipartimento di Matematica, Universit\`a di Bari, I-70125 Bari, Italy}
\affiliation{INFN, Sezione di Bari, I-70126 Bari, Italy}

\author{Rosario Fazio}
\affiliation{International School for Advanced Studies (SISSA),
  Via Beirut 2-4, I-34014 Trieste, Italy}
\affiliation{NEST-CNR-INFM \& Scuola Normale Superiore,
  Piazza dei Cavalieri 7, I-56126 Pisa, Italy}

\author{G. Florio}
\affiliation{Dipartimento di Fisica, Universit\`a di Bari, I-70126 Bari, Italy}
\affiliation{INFN, Sezione di Bari, I-70126 Bari, Italy}

\author{D.A. Lidar}
\affiliation{Departments of Chemistry, Electrical Engineering, and Physics,
  Center for Quantum Information Science \& Technology,
  University of Southern California, Los Angeles, CA 90089}

\author{S. Pascazio}
\affiliation{Dipartimento di Fisica, Universit\`a di Bari, I-70126 Bari, Italy}
\affiliation{INFN, Sezione di Bari, I-70126 Bari, Italy}

\author{F. Plastina}
\affiliation{Dip. Fisica, Universit\`a della Calabria, \&
  INFN-Gruppo collegato di Cosenza, 87036 Arcavacata di Rende (CS), Italy}

\author{P. Zanardi}
\affiliation{Department of Physics,
  Center for Quantum Information Science \& Technology,
  University of Southern California, Los Angeles, CA 90089-0484}
\affiliation{Institute for Scientific Interchange,
  Viale Settimio Severo 65, I-10133 Torino, Italy}

\title{Bang-Bang control of a qubit coupled to a quantum critical spin bath}

\date{\today}

\begin{abstract}

  We analytically and numerically study the effects of pulsed control on the
  decoherence of a qubit coupled to a quantum spin bath. When the environment
  is critical, decoherence is faster and we show that the control is
  relatively more effective. Two coupling models are investigated,
  namely a qubit coupled to a bath via a single link and a spin star model,
  yielding results that are similar and consistent.

\end{abstract}

\pacs{03.65.Yz, 03.67.Pp, 03.67.Hk, 05.70.Jk}
\maketitle


\section{Introduction}

\label{sec:intro}

Decoherence results from the unavoidable coupling between any quantum system
and its environment, and is responsible for the dynamical destruction
of quantum superpositions.
It is detrimental for quantum information
processing~\cite{nielsen} since it leads to a loss of the quantum parallelism
that is implicit in the superposition principle.
The possibility of preventing or avoiding decoherence is hence of significant
importance for any technological use of quantum systems, aimed at
processing, communicating or storing information. To this end, one must
understand and model all of the relevant features characterizing the
environment of the physical system to be protected. Understanding decoherence
is also of fundamental interest in its own right, since it is at the basis of
the description of the quantum-classical transition~\cite{decoherencereview}.

The study of open quantum systems has a long history, and many ingenious
models have been proposed in order to describe the action of the environment
in a quantum dynamical framework (see, e.g., Ref.~\cite{FKM}).
Paradigmatic models represent the environment as a set of harmonic
oscillators~\cite{weiss} or spins~\cite{prokofiev}.
Recently there has been a renewed
interest in the analysis of decoherence induced by such spin
baths~\cite{paganelli,tessieri,khveshchenko,Breuer:04,cucchietti,dawson,
lages05,quan,cucchietti2,rossini07,borz07,zurek07,hamdouni07,camalet07,
alvarez07,Krovi:07,cormick07,relano07,yuan07};
these are clearly relevant in a number of physically important situations,
such as NMR~\cite{decNMR} or spin qubits~\cite{loss}, where loss of
coherence is induced by the coupling to nuclear spins~\cite{decnuclear}.
Several questions have been addressed so far and the picture that emerges
is rather rich. A possible, monogamy-like, relation between
the entanglement in the bath and decoherence has been put forward
in Ref.~\cite{dawson} and subsequently analyzed in different papers.
The signatures of criticality of the environment in decoherence have been
discussed through the study of solvable one-dimensional model
systems~\cite{quan,rossini07,camalet07}. A universal regime exists, in the
strong coupling limit, in which the decay of the Loschmidt echo~\cite{lereview}
does not depend on the system-bath coupling~\cite{cucchietti2,cormick07}.

Several different protocols have been designed to protect quantum
information. These include passive correction techniques, in which
quantum information is encoded in such a way as to suppress the
coupling with the environment~\cite{topological,dfs}, and active
approaches such as quantum error correction~\cite{nielsen} and
dynamical decoupling techniques~\cite{viola98,DD,FLP,theorem} (for
an overview see, e.g.,~\cite{BBreview}, for historical references
see~\cite{history}). Dynamical decoupling strategies aim, by means
of a dynamical control field, at averaging to zero the unwanted
interaction with the environment. In its simplest version, which we
consider here, the control field comprises a train of instantaneous
pulses (\textquotedblleft bang-bang\textquotedblright\ control).
While previous work on dynamical decoupling has made clear
distinctions between different environments, in particular bosonic
baths~\cite{viola98} versus spin baths~\cite{KL:07,Zhang:07}, and
fast versus slow $1/f$ noise~\cite{bangoneoverf}, no attention has
been paid so far to the impact a quantum critical environment might
have on the efficacy of decoupling protocols. This is our goal in
the present work: here we study bang-bang decoupling in the case
where the quantum environment can become critical.

Many-body environments displaying critical behavior have been recently
investigated in great detail, in order to study the sensitivity of decoherence
to environmental dynamics (see, e.g.,~\cite{quan,rossini07}). Close to a
quantum critical point the environment becomes increasingly slower (a
phenomenon known as critical slowing down). We analyze the decoherence
process of a two level system (qubit) coupled to an environment modeled as a
one-dimensional lattice of spins interacting through an Ising-like coupling.
We focus on the suppression of qubit decoherence through a bang-bang control
procedure, and study how the occurrence of a quantum phase transition (QPT)
in the bath modifies the effectiveness of the control procedure.
Our analysis is focused on the behavior of the Loschmidt Echo
(LE)~\cite{lereview}, whose study has given new insights into the decoherence
process of quantum spin chains. We discuss the application of a
pulse train to the qubit and show its effectiveness in quenching qubit
dephasing, especially at the critical point.

The paper is organized as follows. In Sec.~\ref{sec:model} we introduce the
model and pertinent notation. The
control procedure, based on a sequence of pulses that repeatedly flip the
state of the system, is described in Sec.~\ref{sec:control}, where we also
derive an explicit expression for the LE in the presence of such control. We
then provide a detailed analysis of its effects in the limiting cases of a
single qubit-bath link (Subsec.~\ref{sec:controlsingle}) and a spin-star
model (Subsec.~\ref{sec:controlbath}). Finally, in Sec.~\ref{sec:conclusion}
we discuss our results. In the Appendices we provide an analytical formula for
evaluating the LE in the presence of control (App.~\ref{sec:Appendixechofree}),
we perform a perturbative analysis in the pulse frequency of the LE
(App.~\ref{sec: AppendixLE}), and discuss in detail a closed-form formula
for the LE in the spin-star model (App.~\ref{sec:Appendixeffectivedecay}).

\section{Model and notation}

\label{sec:model}

We consider a two level quantum system $S$ (qubit) coupled to an
interacting spin bath $E$ (environment), comprising
a linear chain of $N$ spin-$1/2$ particles, modeled by a transverse
field Ising model. The Hamiltonian reads
\begin{equation}
  \mathcal{H}_{0}=\mathcal{H}_{S}+\mathcal{H}_{E}+\mathcal{H}_{\mathrm{int}}\,,
  \label{eq:HamFree}
\end{equation}
where $\mathcal{H}_{S}$ and $\mathcal{H}_{E}$ are the free Hamiltonians of
$S$ and $E$:
\begin{eqnarray}
  \mathcal{H}_{S} &=&-\frac{\omega _{0}}{2}\,(\mathbbm{1}-\tau ^{z})=-\omega
  _{0}\left\vert \downarrow \right\rangle \left\langle \downarrow \right\vert
  \,, \\
  \mathcal{H}_{E} &=&-J\sum_{j=1}^{N}\Big(\sigma _{j}^{x}\sigma
  _{j+1}^{x}+\lambda \sigma _{j}^{z}\Big)\,;  \label{eq:systemhamilt}
\end{eqnarray}
here $\sigma _{i}^{\alpha }$ and $\tau ^{\alpha }$ (with $\alpha =x,y,z$)
indicate, respectively, the Pauli matrices of the $i$th spin of the chain $E$
and of the qubit $S$, whose ground and excited states are denoted by
$\left\vert \uparrow \right\rangle $ and $\left\vert \downarrow \right\rangle$.
In this work we will use periodic boundary conditions, therefore
we assume $\sigma _{N+1}^{\alpha}\equiv \sigma _{1}^{\alpha }$.
The constants $J$ and $\lambda $ are the
interaction strength between neighboring spins of the bath and an external
transverse magnetic field, respectively (in the following, the energy and
the time scale are taken in units of $J$, therefore, when not specified,
we will implicitly assume $J=1$). We suppose that the system is
coupled to a given number of bath spins~\cite{rossini07}:
\begin{equation}
  \mathcal{H}_{\mathrm{int}}=-\epsilon \,\left\vert \downarrow \right\rangle
  \left\langle \downarrow \right\vert \otimes \sum_{j=j_{1}}^{j_{m}}\sigma
  _{j}^{z}\,,  \label{eq:interactionhamilt}
\end{equation}
where $\epsilon $ is the coupling constant and $m$ the number of
environmental spins to which $S$ is coupled. The LE can be calculated for a
generic sequence $\{j_{1},\ldots j_{m}\}$ of system-bath links.
In the following, however, we consider the cases $m=1$ and $m=N$.
We expect that the generic case will be a quantitative interpolation between
these two extremes but no new qualitative features should emerge.

With the above choice of $\mathcal{H}_S$ and $\mathcal{H}_{\mathrm{int}}$,
the populations of the ground and excited state of the qubit do not evolve,
since $[\tau^z, \mathcal{H}_0 ] =0$, and we can study a model of pure
dephasing.

As usual, we assume that the initial global state of the system is factorized:
\begin{equation}
  \ket{\Psi (0)} = \left( c_{\uparrow} \ket{\uparrow} +
  c_{\downarrow} \ket{\downarrow} \right) \otimes \ket{G} \, ,
  \label{eq:InitialState}
\end{equation}
so that the qubit $S$ is in a generic superposition of the ground
and excited state, while the bath $E$ is in its ground state (i.e.,
$\ket{G}$ is the ground state of the Hamiltonian $\mathcal{H}_{E}$).
The evolution of such a state under the Hamiltonian
\eqref{eq:HamFree} is dictated by the unitary operator $U_{0} =
e^{-i\mathcal{H}_{0}t}$ and yields, at time $t$, the state
\begin{equation}
  \ket{\Psi (t)} = c_\uparrow \ket{\uparrow} \ket{\varphi_{0}(t)}
  + c_\downarrow \, e^{i\omega_{0}t} \ket{\downarrow} \ket{\varphi_{1}(t)} \, ,
\end{equation}
where $\ket{\varphi_{0}(t)} \equiv e^{-i \Ham_\uparrow t} \ket{G}$ and
$\ket{\varphi_{1}(t)} \equiv e^{-i \Ham_\downarrow t} \ket{G}$
are the environment states evolved under an ``unperturbed'' and a
``perturbed'' Hamiltonian given, respectively, by
\begin{equation}
  \mathcal{H}_{\uparrow }\equiv \mathcal{H}_{E} \, , \qquad
  \mathcal{H}_{\downarrow} \equiv \mathcal{H}_{E} +
  \left\langle \downarrow \right\vert \mathcal{H}_{\mathrm{int}}
  \left\vert \downarrow \right\rangle \, .
  \label{eq:Hud}
\end{equation}

The density matrix of the qubit is
$\rho =\mathrm{Tr}_{E}\left\vert \Psi \right\rangle \left\langle \Psi
\right\vert $. Its diagonal elements are constant, while off-diagonal
elements decay in time as
\begin{equation}
  \rho _{\downarrow \uparrow }(t)=\rho _{\downarrow \uparrow }(0)\,e^{i\omega
    _{0}t}D(t),  \label{eq:offred}
\end{equation}
with
\begin{equation}
  D(t)=\langle {\varphi _{0}(t)}|\varphi _{1}(t)\rangle =\left\langle G\right\vert
  e^{i\mathcal{H}_{\uparrow }t}e^{-i\mathcal{H}_{\downarrow }t}\left\vert
  G\right\rangle .  \label{eq:D}
\end{equation}
The decoherence of the qubit is then fully characterized by the so called
\textit{Loschmidt echo} $\mathcal{L}_{0}(t)\in \lbrack 0,1]$
of the environment:
\begin{equation}
  \mathcal{L}_{0}(t)\equiv |D(t)|^{2}= |\left\langle G\right\vert
  e^{-i(\mathcal{H}_{E}+\left\langle \downarrow \right\vert \mathcal{H}_{\mathrm{int}}
    \left\vert \downarrow \right\rangle )t}\left\vert G\right\rangle |^{2} \, .
  \label{eq:LEchoFree}
\end{equation}

The decay of the LE in the
model~\eqref{eq:HamFree}-\eqref{eq:interactionhamilt} with $m=N$
(spin-star model) was first studied in detail in Ref.~\cite{quan};
an extension to the more general case $m\neq N $, and for other spin
baths -- including the $XY$ and Heisenberg models -- can be found in
Ref.~\cite{rossini07}. It was pointed out that the echo decay is
enhanced at criticality, due to the hypersensitivity to
perturbations of the (time-evolved) unperturbed ground state
$\left\vert \varphi _{0}(t)\right\rangle$. Indeed, at criticality
the perturbation $\mathcal{H}_{\mathrm{int}}$ is very effective at
making the unperturbed state $\ket{\varphi_{0}(t)}$ orthogonal to
$\ket{\varphi_{1}(t)}$, thus leading to a strong decay of the echo.
Away from criticality, the perturbation is not so effective at
orthogonalizing $\ket{\varphi _{0}(t)}$ and $\ket{\varphi_{1}(t)}$,
whence the echo decays more slowly. In the following we investigate
these effects when a control is also present. Details on how to
evaluate the LE in both the absence and presence of such a control
are given in Appendix~\ref{sec:Appendixechofree}.

\section{Controlled dynamics}

\label{sec:control}

Quantum dynamical decoupling procedures aimed at actively fighting decoherence
hinge either on the action of frequent interruptions of the evolution or on the
effect of a strong continuous coupling to an external field. These
procedures are known to be physically and, to a large extent, mathematically
equivalent~\cite{FLP}.
Here we focus on one possible procedure, based on multipulse
control~\cite{viola98}.
Let us formally introduce the control scheme as
\begin{equation}
  \mathcal{H}(\omega _{0},t)=\mathcal{H}_{0}+\mathcal{H}_{P}(\omega _{0},t)\,,
  \label{eq:HamGlobal}
\end{equation}
where $\mathcal{H}_{P}$ is an additional time-dependent Hamiltonian that
causes spin flips of the qubit at regular time intervals through a
monochromatic alternating magnetic field at resonance:
\begin{eqnarray}
  \mathcal{H}_{P}(\omega _{0},t) & = & \sum_{n} V^{(n)}(t)
  \Big[\cos \big( \omega_{0}(t-n\Delta t)\big) \,\tau _{x}  \notag \\
    & & + \sin \big(\omega _{0}(t-n\Delta t)\big) \,\tau _{y} \Big] \,.
\end{eqnarray}
Here $V^{(n)}(t)$ is constant and equal to $V$ for the entire duration $\tau_{P}$
of the $n$th pulse (i.e., for $n\Delta t\leq t\leq n\Delta t+\tau _{P}$),
$\Delta t$ being the time interval between two consecutive pulses.
In this work we only deal with $\pi$ pulses, satisfying the condition
$2 V \tau_P = \pm \pi$, and suppose that $V$ is large enough to yield almost
instantaneous spin flips, i.e., we take $\tau_P \ll \Delta t$.
Therefore, in the ideal limit of instantaneous kicks of infinite strength
($\tau_P \to 0, \, V \to \infty$ such that $V \tau_P = \pm \pi / 2$),
the effect of each pulse on the qubit is simply a flip, that is described
by the operator
\begin{equation}
  U_P = \pm i \, \tau^x.
\end{equation}

The evolution of the initial state~\eqref{eq:InitialState} under the
Hamiltonian~\eqref{eq:HamGlobal} in one spin-flip cycle [i.e., two flips,
from time $t=0$ to time $t_{1}=2(\Delta t+\tau _{P})\simeq 2\,\Delta t$] is
dictated by the unitary operator
\begin{equation}
  U_{\mathcal{C}}\equiv e^{-2i\mathcal{H}\Delta t}=U_{P}\;U_{0}(\Delta t)
  \; U_{P} \; U_{0}(\Delta t)
\end{equation}
and it is such that
\begin{eqnarray}
  \ket{\Psi (2 \Delta t)} & = & - c_\uparrow  e^{- i \omega_0 \Delta t} \ket{\uparrow}
  e^{- i \Ham_\downarrow \Delta t} e^{- i \Ham_\uparrow \Delta t} \ket{G} \nonumber \\
  & & - c_\downarrow e^{- i \omega_0 \Delta t} \ket{\downarrow}
  e^{- i \Ham_\uparrow \Delta t} e^{- i \Ham_\downarrow \Delta t} \ket{G} .\nonumber \\
\end{eqnarray}
This is again a pure dephasing phenomenon, so that all relevant information
is contained in the off-diagonal element~\eqref{eq:offred} of the system reduced
density matrix. The behavior of decoherence is then fully captured by the LE:
\begin{equation}
  \mathcal{L}(2\Delta t) = \Big\vert \left\langle G\right\vert
  \left( e^{i \mathcal{H}_{\downarrow} \Delta t}
  e^{i\mathcal{H}_{\uparrow }\Delta t}\right) \cdot
  \left( e^{-i\mathcal{H}_{\downarrow }\Delta t} e^{-i\mathcal{H}_{\uparrow }\Delta t}\right)
  \left\vert G\right\rangle \Big\vert^{2}\,.
\end{equation}
In general, at a certain time $t=2M \Delta t+\tilde{t}$, the evolution
operator of the global system is given by:
\begin{equation}
  U=\left\{
  \begin{array}{ll}
    U_{0}(\tilde{t}) \: [U_{\mathcal{C}}]^{M} & \mathrm{if} \quad
    \tilde{t} < \Delta t \vspace{1mm} \\
    U_{0}(\tilde{t}-\Delta t) \, U_{P} \, U_{0}(\Delta t) \: [U_{\mathcal{C}}]^{M}
    & \mathrm{if} \quad \tilde{t} \geq \Delta t
  \end{array}
  \right.
\end{equation}
where $M=[\frac{t}{2\Delta t}]$, $[\cdot ]$ denotes the integer part and
$\tilde{t}\equiv t- 2M \Delta t$ is the residual time after $M$ cycles.
It is now easy to write down the LE at a generic time t:
\begin{widetext}
  \begin{equation}
    \mathcal{L} (t) = \left\{ \begin{array}{ll}
      \Big\vert \, \langle G \vert \left( e^{i \Ham_\downarrow \Delta t}
      e^{i \Ham_\uparrow \Delta t} \right)^M e^{i \Ham_\downarrow \tilde{t}}
      \cdot e^{- i \Ham_\uparrow \tilde{t}} \left( e^{-i \Ham_\downarrow \Delta t}
      e^{-i \Ham_\uparrow \Delta t} \right)^M \ket{G} \, \Big\vert^2 & \quad
      \mathrm{if} \;\; \tilde{t} < \Delta t  \\
      \Big\vert \, \langle G \vert \left( e^{i \Ham_\downarrow \Delta t}
      e^{i \Ham_\uparrow \Delta t} \right)^M e^{i \Ham_\downarrow \Delta t}
      e^{i \Ham_\uparrow \tilde{t}} \cdot e^{- i \Ham_\downarrow \tilde{t}}
      e^{- i \Ham_\uparrow \Delta t} \left( e^{-i \Ham_\downarrow \Delta t}
      e^{-i \Ham_\uparrow \Delta t} \right)^M \ket{G} \, \Big\vert^2 & \quad
      \mathrm{if} \;\; \tilde{t} \geq \Delta t \end{array} \right.
    \label{eq:LEchoPulse}
  \end{equation}
\end{widetext}
An explicit formula for evaluating the LE, also in the presence of pulses,
is given in Appendix~\ref{sec:Appendixechofree}.

In the limit of short pulse intervals, and when $t$ is an integer
multiple of the duration of a single spin-flip cycle, $t=2 M\Delta
t$, one can show (see Appendix~\ref{sec: AppendixLE}) that
Eq.~\eqref{eq:LEchoPulse} can be rewritten as
\begin{equation}
  \mathcal{L}(t = 2 M\Delta t)=\Big\vert \left\langle G \right\vert
  e^{i t  \mathcal{H}_{\mathrm{eff}}} \left\vert G\right\rangle
  \Big\vert^2 + M \, O(\Delta t^{2}) \, ,
  \label{eq:LE-eff}
\end{equation}
where
\begin{equation}
  \mathcal{H}_{\mathrm{eff}}\equiv i\frac{\Delta t}{2}[\mathcal{H}_{\downarrow },
    \mathcal{H}_{\uparrow }]=i\frac{\Delta t}{2}\,[\left\langle \downarrow
    \right\vert \mathcal{H}_{\mathrm{int}}\left\vert \downarrow \right\rangle
    ,\mathcal{H}_{E}]
\label{eq:H-eff}
\end{equation}
is an effective Hamiltonian. By noting that $\left\langle \downarrow
\right\vert \mathcal{H}_{\mathrm{int}}\left\vert \downarrow \right\rangle
=-\epsilon \sum_{j=j_{1}}^{j_{m}}\sigma _{j}^{z}$, we have, for arbitrary $\lambda$
\begin{equation}
  \mathcal{H}_{\mathrm{eff}} = i\epsilon _{\mathrm{eff}} \bigg[
    \sum_{j=j_{1}}^{j_{m}}\sigma _{j}^{z},\sum_{j=1}^{N}\sigma _{j}^{x}\sigma
    _{j+1}^{x}\bigg] \, ,
  \label{eq:Heff}
\end{equation}
where
\begin{equation}
  \epsilon _{\mathrm{eff}}\equiv \epsilon J\frac{\Delta t}{2}.  \label{eq:renormcc}
\end{equation}
This is the renormalized system-bath coupling constant in the presence of
multipulse control. We notice that $\mathcal{H}_{\mathrm{eff}}$ does not
depend on $\lambda$ (which would appear at $O(\Delta t^{3})$ through the
double commutator
$[[\mathcal{H}_{\downarrow },\mathcal{H}_{\uparrow }], \mathcal{H}_{\uparrow }]$).
Therefore, in the small $\Delta t$ limit, the criticality of the model
can manifest itself only through $\ket{G}$ in the LE expression~\eqref{eq:LE-eff}.

In the next two subsections we turn to a numerical study of the LE for the
cases of a qubit coupled to one spin of the chain
[$m=1$ in Eq.~\eqref{eq:interactionhamilt}], and the spin-star model
[$m=N$ in Eq.~\eqref{eq:interactionhamilt}].

\subsection{Qubit coupled to a single bath spin}

\label{sec:controlsingle}

When $m=1$, the system-bath Hamiltonian of
Eqs.~\eqref{eq:HamFree}-~\eqref{eq:interactionhamilt} can be
rewritten as:
\begin{equation}
  \mathcal{H}_{0}=-\left\vert \downarrow \right\rangle \left\langle \downarrow
  \right\vert \big(\omega _{0}+\epsilon \,\sigma _{1}^{z}\big)-J\sum_{j=1}^{N}
  \Big( \sigma _{j}^{x}\sigma _{j+1}^{x}+\lambda \sigma _{j}^{z}\Big)
  \label{eq:onespin}
\end{equation}
and corresponds to a situation in which the qubit is directly coupled to
only one spin of an Ising chain with periodic boundary conditions
(the coupled bath-spin qubit is assumed for simplicity and with no
loss of generality to be the first one).
%
\begin{figure}[!ht]

\includegraphics[width=0.45 \textwidth]{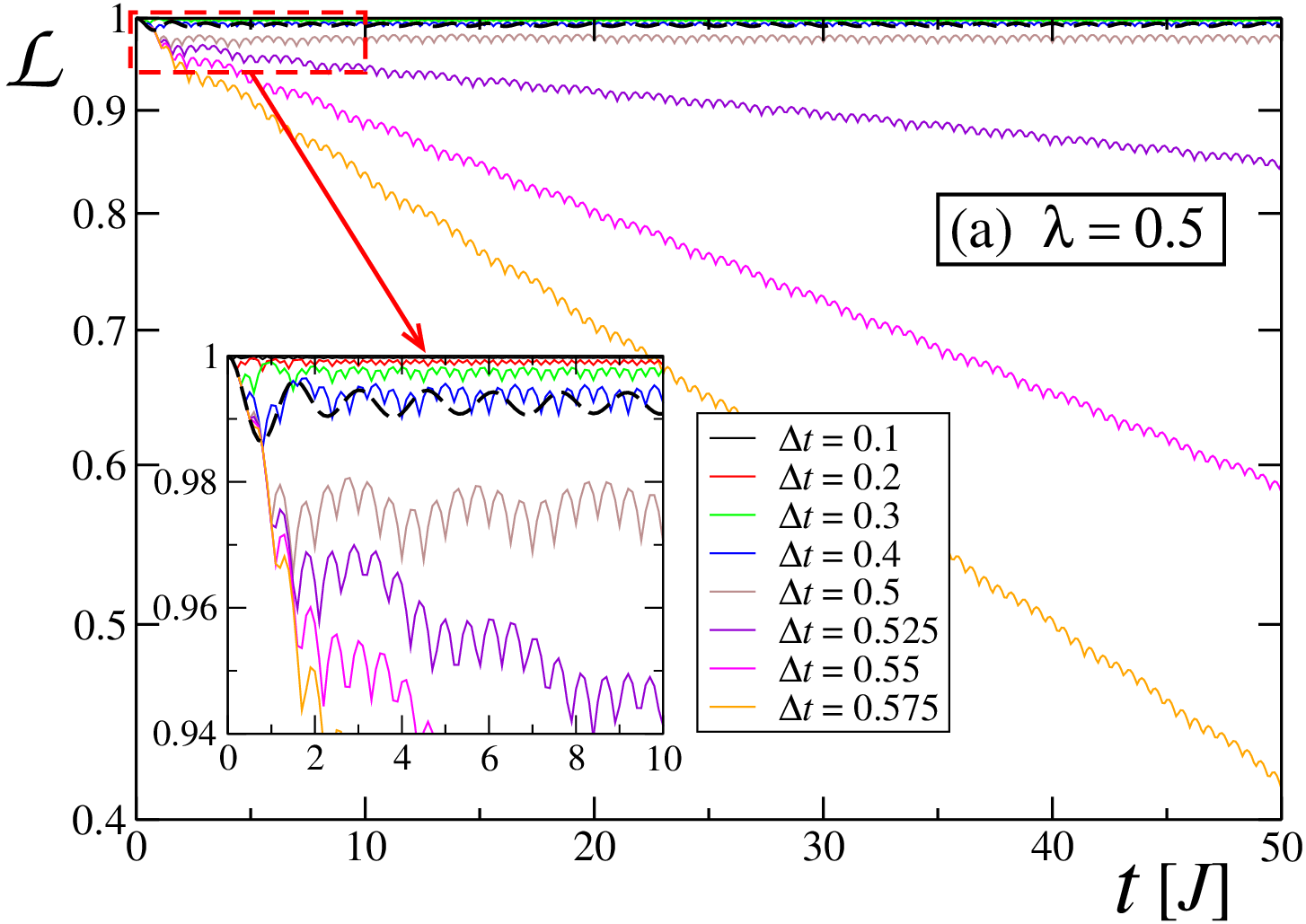}
\includegraphics[width=0.45 \textwidth]{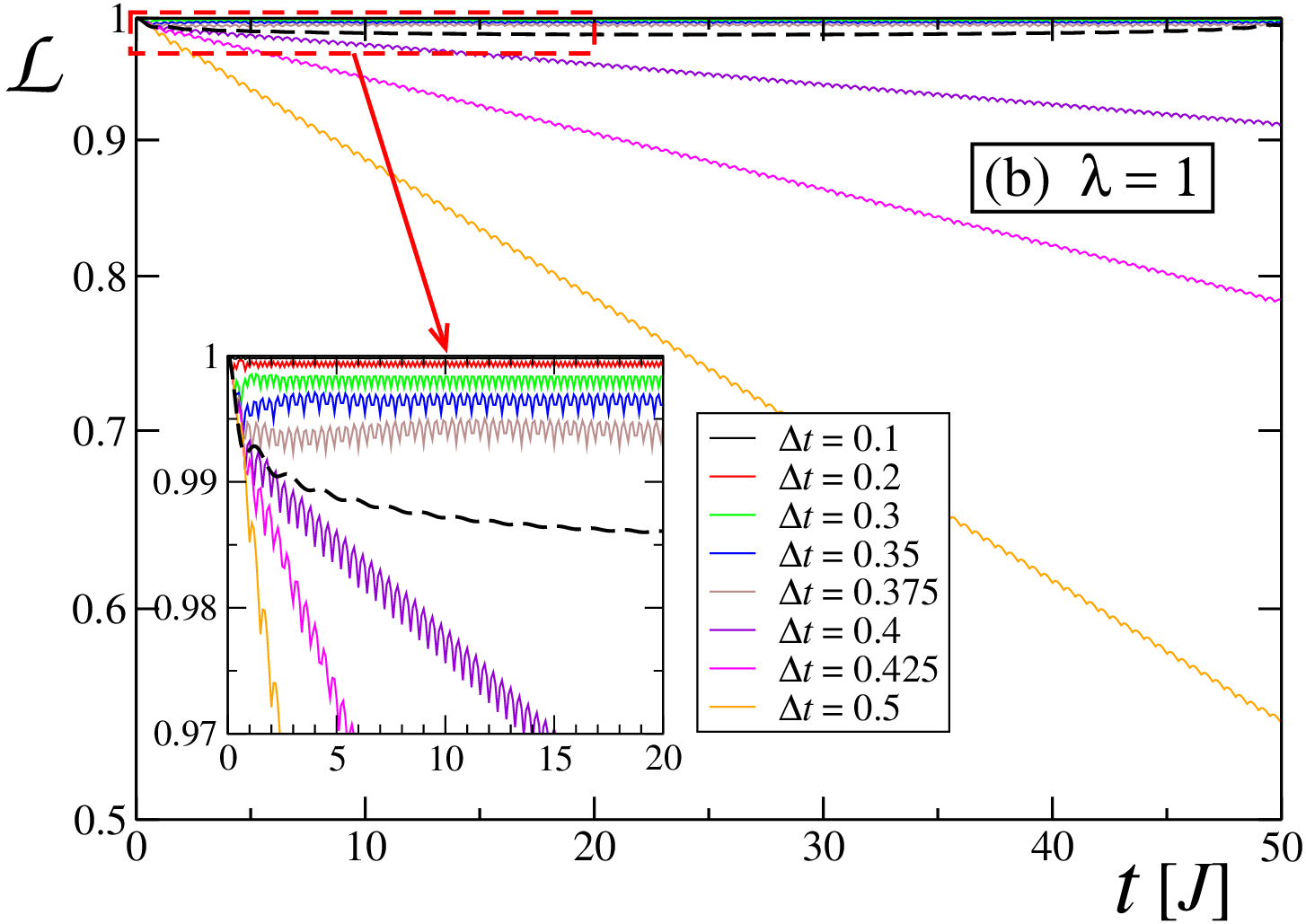}
\includegraphics[width=0.45 \textwidth]{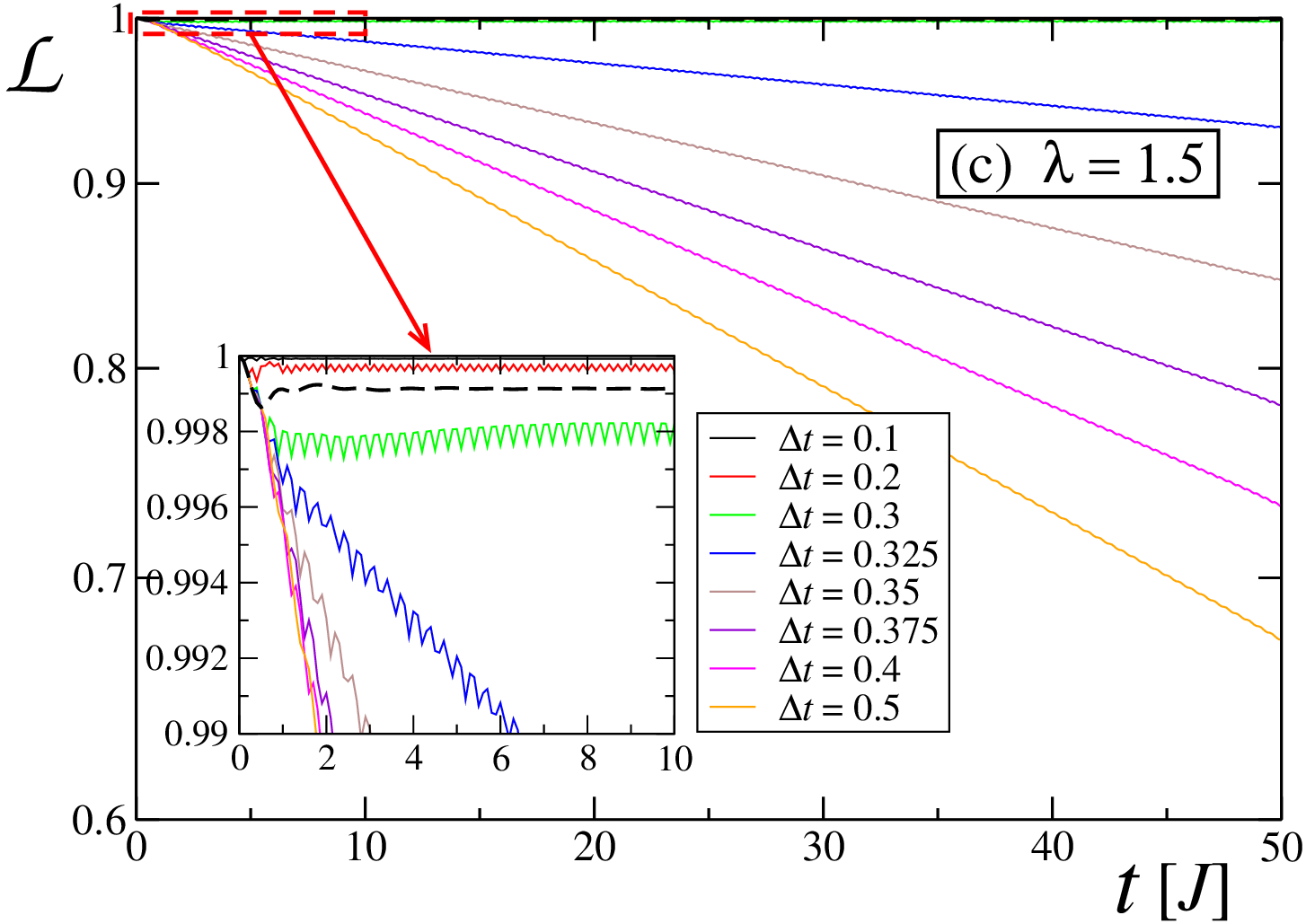}
\caption{(Color online) Loschmidt Echo as a function of time for a qubit
  coupled to a $N=100$ spin Ising chain, with $\epsilon = 0.25$.
  Panels stand for different values of the transverse field:
  (a) $\lambda = 0.5$, (b)  $\lambda = 1$, (c)  $\lambda = 1.5$;
  the various curves in each panel are for decreasing pulse intervals
  $\Delta t$, from top to bottom.
  Insets: magnification at small times $t$ (axes units are the same
  as in main panels); notice that, when $\protect\lambda=1$,
  frequent pulses suppress decay for $J \Delta t \lesssim 0.375$
  (here and in the following figures $\Delta t$ values are expressed
  in units of $J$).}
\label{fig:lecho_time}
\end{figure}
%
In Fig.~\ref{fig:lecho_time} we show the behavior of the LE in
Eq.~\eqref{eq:LEchoPulse} as a function of time, for different values of the
pulse frequency $\Delta t$.
The three panels refer to different values of the transverse magnetic field
$\lambda $; the thick dashed lines represent the case $\mathcal{L}_{0}(t)$
with no external control [$\Delta t\rightarrow \infty$ in
Eq.~\eqref{eq:LEchoPulse}, or simply Eq.~\eqref{eq:LEchoFree}].
Here the environment consists of $N=100$ Ising spins, and the system-bath
coupling has been set at $\epsilon =0.25$.

We notice a very different behavior as $\lambda $ is varied. Away
from criticality (i.e., for $\lambda =0.5$ Fig.~\ref{fig:lecho_time}(a),
and $\lambda = 1.5$ Fig.~\ref{fig:lecho_time}(c))
the LE in absence of control quickly reaches its
asymptotic (saturation) value $\mathcal{L}_{\infty}$, as indicated
by the dashed black lines. Very fast control pulses do improve the
situation, but only in the sense that this asymptotic value becomes
slightly closer to unity. In contrast, slow pulses make the
situation much worse: when $J \Delta t$ is larger than a certain
value, the pulses act as an additional source of noise and, as a
consequence, the coherence decays (exponentially). On the other
hand, when the chain is critical ($\lambda =1$ Fig.~\ref{fig:lecho_time}(b))
and there is no control, the LE decays (albeit only
logarithmically~\cite{rossini07}), as can be seen from the dashed
curve. In this case the pulses can be very
effective, as a control procedure: when $J \Delta t\lesssim 0.375$
decay is suppressed. Again, when $\Delta t$ exceeds this threshold,
decay is enhanced. This situation is reminiscent of the transition
between a quantum Zeno and an inverse Zeno effect~\cite{invZeno}.

\begin{figure}[!t]
\includegraphics[width=0.45 \textwidth]{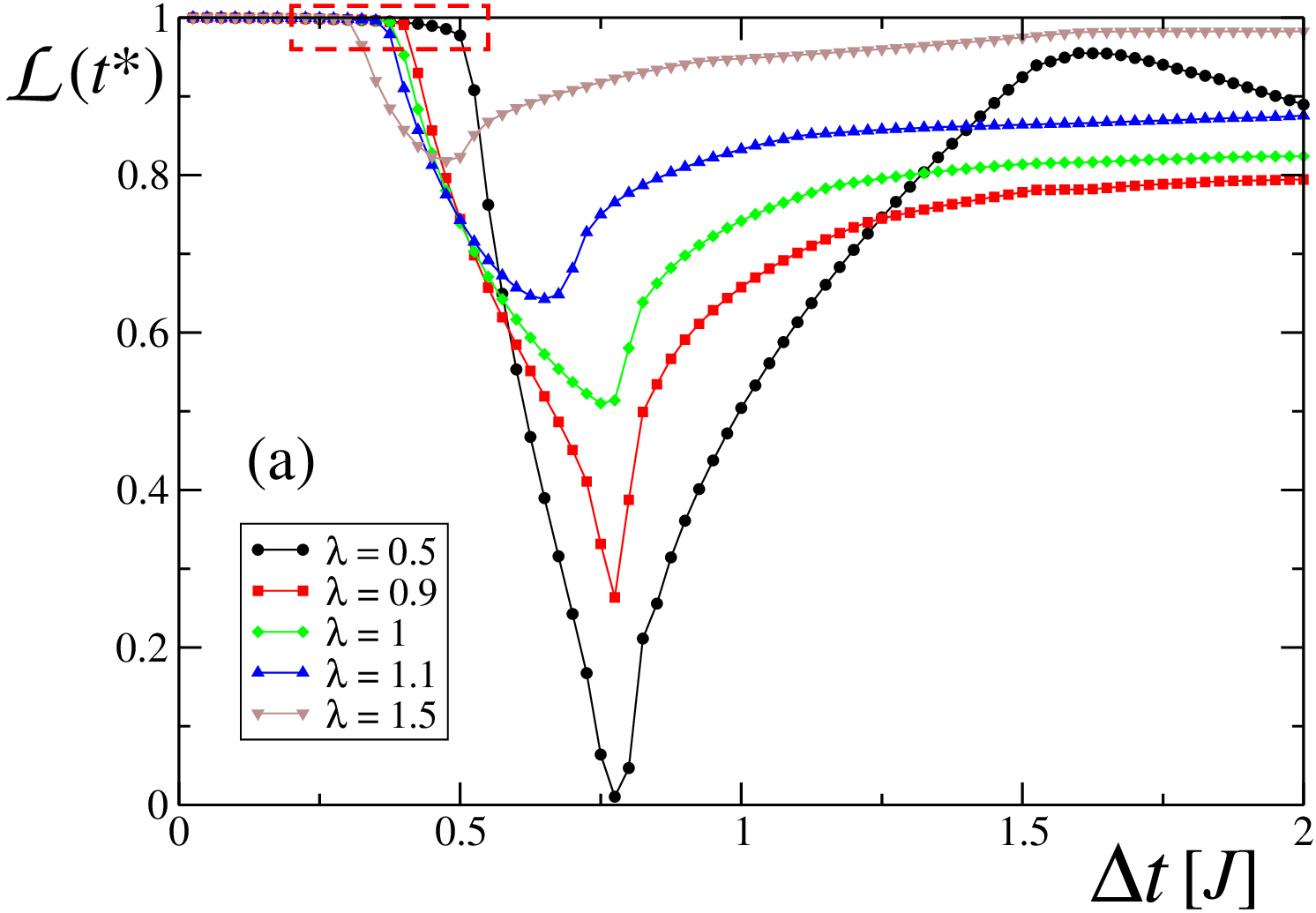}
\includegraphics[width=0.45 \textwidth]{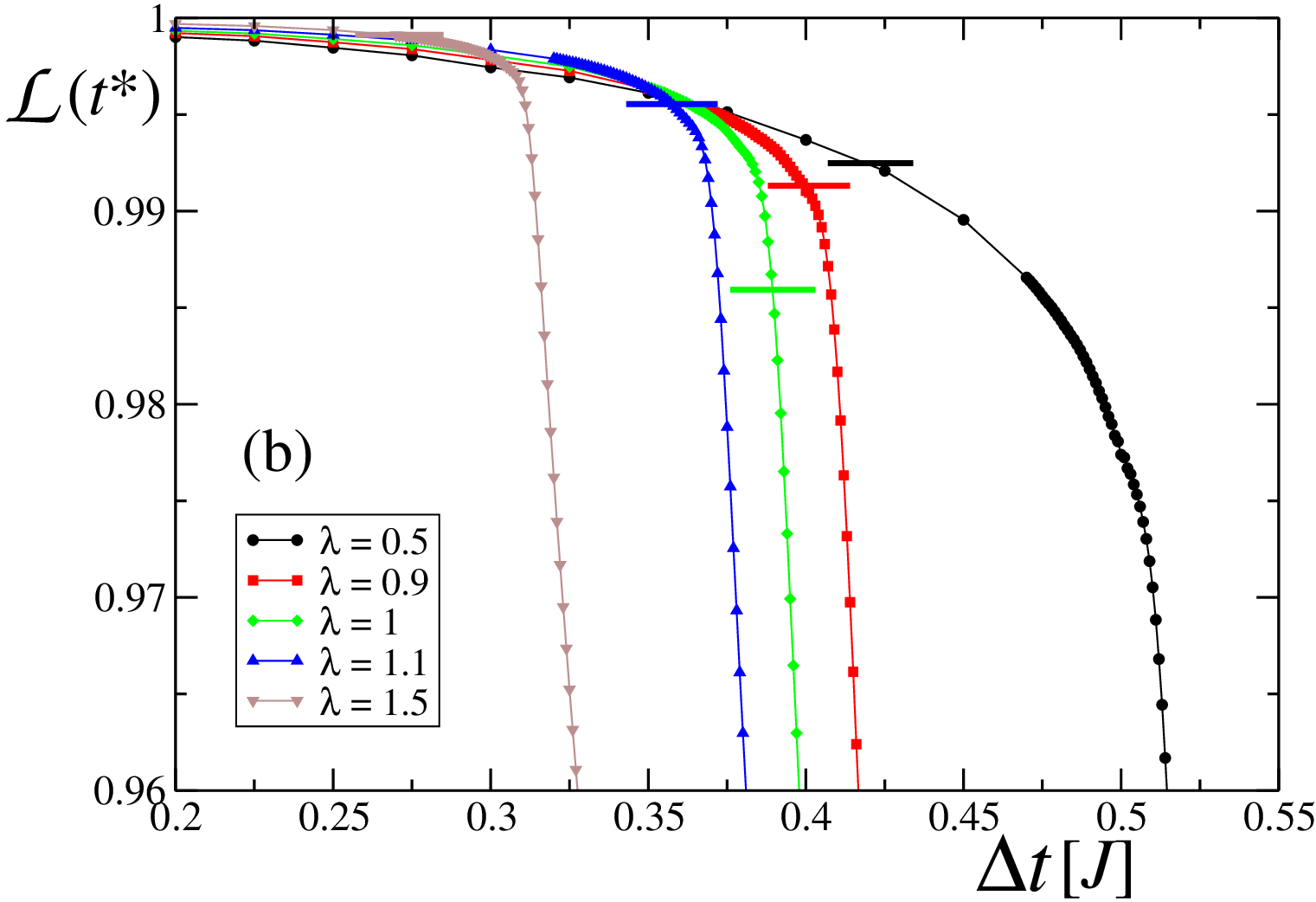}
\caption{(Color online) (a) LE as a function of the pulse frequency
  $\Delta t$ at a given time $t^*$, for different $\protect\lambda$.
  (b) Magnification of panel (a) in the highlighted zone;
  the bars denote the corresponding values of $\mathcal{L}_0 (t^*)$ without
  pulsing. Here we set $J t^*=25$, $N=100$, $\protect\epsilon = 0.25$.}
\label{fig:plat_dt}
\end{figure}

In Fig.~\ref{fig:plat_dt} we show the values of the LE at a fixed
time $t^{\ast}$ (we performed an average of $\mathcal{L}(t)$ for
$Jt\in \lbrack Jt^{\ast }-5,Jt^{\ast }+5]$ in order to eliminate fast
oscillations), as a function of $\Delta
t$. The different curves are obtained for different values of the
transverse field $\lambda $. We set $Jt^{\ast }=25$ so that: i) in
the absence of pulse control and for noncritical $\lambda $,
$\mathcal{L}_{0}$ has already reached its saturation value
$\mathcal{L}_{\infty }$; ii) at criticality, the minimum of
$\mathcal{L}_{0}(t)$ is found exactly at $Jt^{\ast }\simeq N/4$ (in
this case $N=100$)~\cite{rossini07}.
In the panel (b), bars denote the corresponding value of
$\mathcal{L}_{0}(t^{\ast })$ without external control.

The behavior at large pulse intervals $\Delta t$ is non-trivial and rather
interesting:\ we note that the echo has a minimum and has
an almost complete recovery, and
that the LE for $\lambda=0.5$ rises higher than for $\lambda=0.9,1,1.1$.
The large $\Delta t$ regime is non-perturbative (in the sense of the
perturbation theory of Section~\ref{sec:control} and
Appendix~\ref{sec: AppendixLE}).
Nevertheless, the rise of the LE for large $\Delta t$ can be understood
as being due to the fact that, after a short transient time $\bar{t}$,
the LE \emph{without control} saturates around a constant value
(see the black dashed curves in the insets of Fig.~\ref{fig:lecho_time},
or Ref.~\cite{rossini07}).
Therefore, if the pulse frequency is such that $\Delta t > \bar{t}$,
the effect of the bang-bang control procedure will be progressively reduced
as $\Delta t$ grows, until, in the limit $\Delta t \to + \infty$, it
will completely disappear.
In other words, the detrimental effect of the control for large $\Delta$
is offset by the gradual diminishing of its effect as $\Delta t$ grows,
which allows the LE to recover to its saturation value.
Moreover, as the insets of Fig.~\ref{fig:lecho_time} show, for $\lambda=1.5$
the saturation is truly at a constant value; for $\lambda=0.5$ the saturation
is an oscillation around a constant value; at criticality ($\lambda=1$)
there is a logarithmic decay of the LE, but for a finite system size
this decay will eventually stop and revivals of quantum coherence will appear.
The oscillation at $\lambda = 0.5$ explains why this curve rises higher than
the other curves in Fig.~\ref{fig:plat_dt}(a); at a time $t^* \approx 1.5$,
the uncontrolled LE in Fig.~\ref{fig:lecho_time} at $\lambda = 0.5$
is larger than for other values of $\lambda$.

\begin{figure}[!t]
\includegraphics[width=0.45 \textwidth]{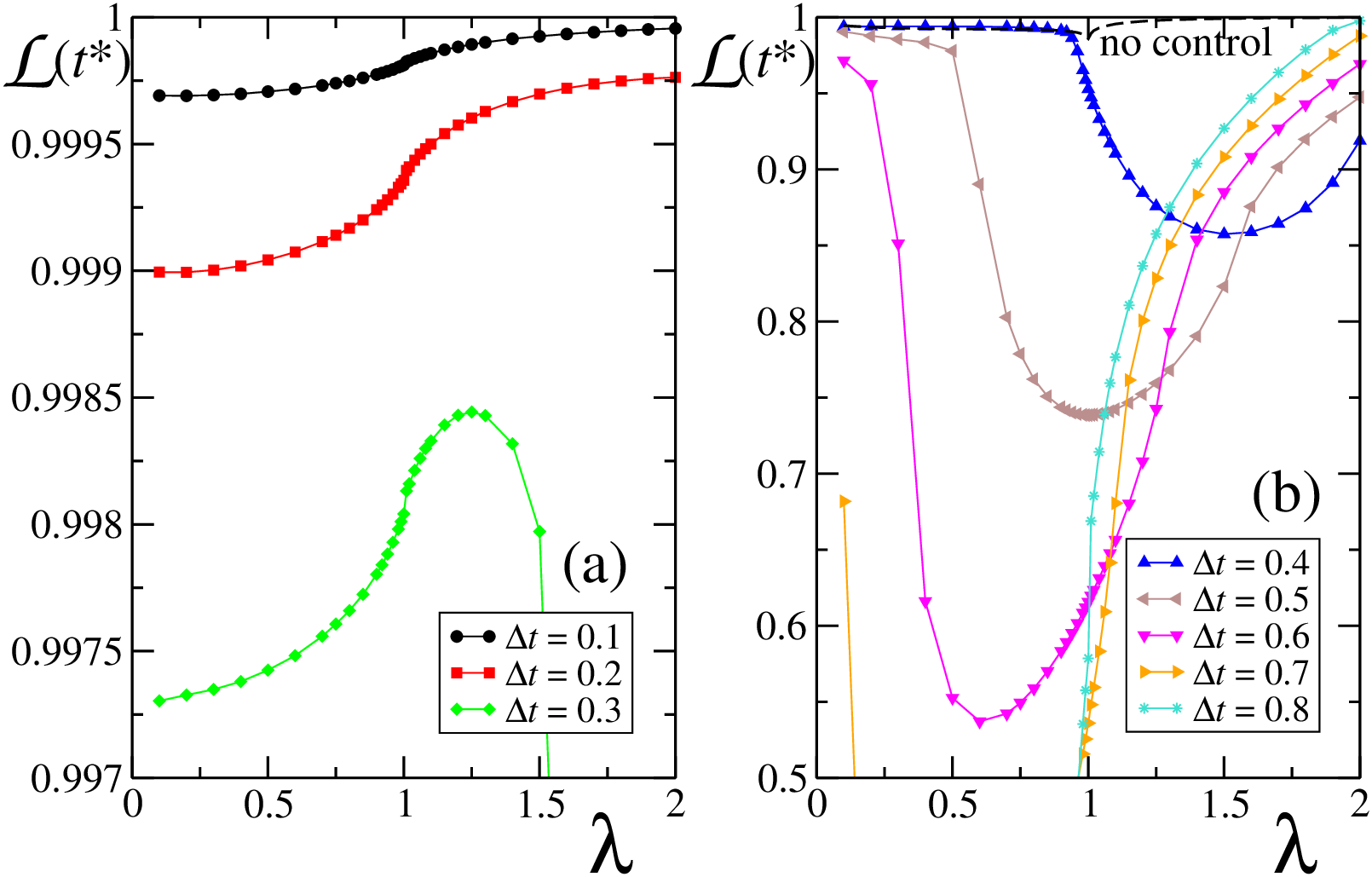}
\includegraphics[width=0.45 \textwidth]{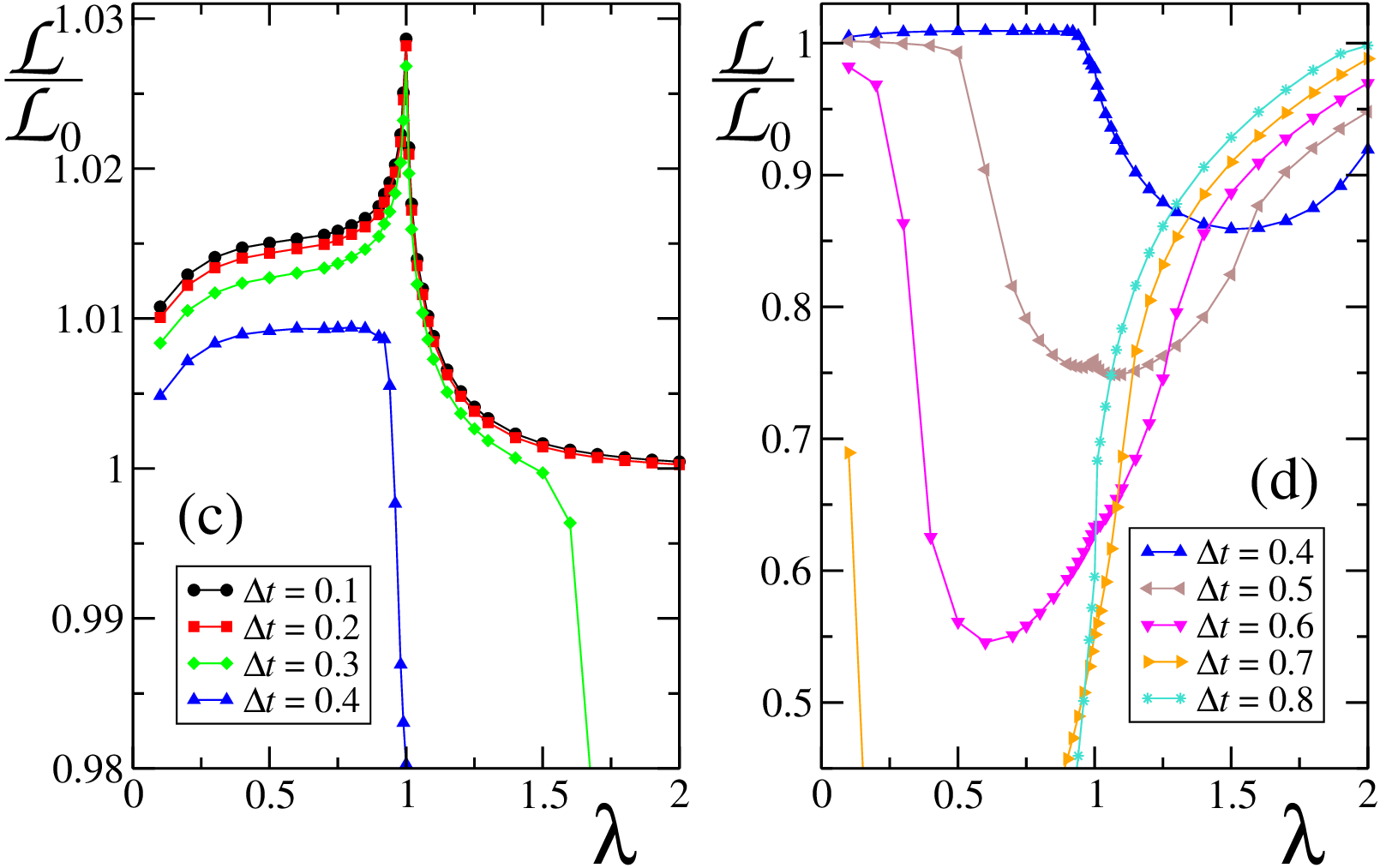}
\caption{(Color online) Panels (a)-(b): LE at a fixed time $t^{\ast }$ as a
  function of the transverse field, for different values of $\Delta t$.
  Panels (c)-(d): rescaled LE, $\mathcal{L}(t^{\ast })/\mathcal{L}_{0}(t^{\ast })$.
  Notice the widely different scales in the $y$ axes of (a)-(c) panels
  (small $\Delta t$), with respect to (b)-(d) panels (large $\Delta t$).
  Here we set $J t^{\ast}=25 $, $N=100$, $\protect\epsilon =0.25$.}
\label{fig:frac}
\end{figure}

The panels (a)-(b) of Fig.~\ref{fig:frac} display $\mathcal{L}(t^{\ast})$
as a function of $\lambda $, for different values of $\Delta t$.
In panels (c)-(d) we plot the rescaled quantities, obtained by
dividing $\mathcal{L}(t^{\ast })$ by the corresponding value in
absence of pulse control, $\mathcal{L}_{0}(t^{\ast})$. The LE has a
maximum not at $\lambda =1$ but at $\lambda >1$, while at $\lambda
=1$ there is an inflexion point. At criticality, the rescaled LE
displays a cusp. The cusp disappears at $J \Delta t\gtrsim 0.375$, in
agreement with Fig.~\ref{fig:lecho_time}(b), where
we observed, at the same value of $\Delta t$, an increase of the LE
when the control is present. A qualitative explanation of this
phenomenon is straightforward: for short time pulses, the
renormalized coupling constant $\epsilon_{\mathrm{eff}}$ in
Eq.~\eqref{eq:renormcc}, and therefore the LE, are only weakly
dependent on $\lambda$ at leading order in the perturbative
expansion. In contrast, the free echo $\mathcal{L}_{0}$ has a
downward cusp~\cite{rossini07} (present also in Fig.~\ref{fig:SpinStar}(a)
for the spin-star case). The ratio must
therefore display an upward cusp, as seen in Fig.~\ref{fig:frac}.
Another way to state this explanation is the following. For
sufficiently small values of $\Delta t$ the bang-bang protocol
succeeds at effectively eliminating the environment action. The only
remnant of criticality is then the weak signature of an inflexion
point seen in Fig.~\ref{fig:frac}(a). The echo
of the uncontrolled system, however, is hypersensitive to
criticality, as indicated by the cusp. On the other hand, when
$\Delta t$ is too large (Fig.~\ref{fig:frac}(b)-(d)), the
bang-bang protocol fails at removing the coupling of the qubit to
the environment, and the controlled and uncontrolled echos behave
similarly.

There are other interesting features in Fig.~\ref{fig:frac}.
Panels (a)-(b) show that the LE rises for sufficiently large
$\lambda$, and (c)-(d) show that the ratio between the
decoupled and free echoes approaches unity for large $\lambda$. This
can be understood as being due to the dominance of the uniform
magnetic field term $\lambda \sum_{j=1}^{N}\sigma _{j}^{z}$ over the
transverse Ising term $\sum_{j=1}^{N}\sigma _{j}^{x}\sigma
_{j+1}^{x}$ in Eq.~\eqref{eq:onespin}. Indeed, in the limit of large
$\lambda$, this means that $\mathcal{H}_{\downarrow }\simeq
\mathcal{H}_{E}$ [recall Eq.~\eqref{eq:Hud}], so that
$[\mathcal{H}_{\uparrow }, \mathcal{H}_{\downarrow}]\simeq 0$ and
the LE $\simeq 1$ by Eqs.~\eqref{eq:D} and~\eqref{eq:LEchoFree}.
Thus, at large $\lambda$, decoupling is not needed to obtain a large LE.

More interesting is the monotonic rise of the LE visible in panel (a)
as a function of $\lambda $ for $J \Delta t=0.1,0.2$, in contrast to
the maximum around $\lambda \sim 1.25$ for $J \Delta t=0.3$.
Indeed, panel (c) shows that decoupling makes the situation worse
for $J \Delta t=0.3$ and $\lambda \gtrsim 1.25$, and a similar trend
continues in panels (b)-(d). Thus, in our model decoupling is fully effective
(i.e., for all values of $\lambda $) for $J \Delta t\lesssim 0.2$.

\subsection{Spin-star model}

\label{sec:controlbath}

The ``spin-star'' model corresponds to the case when the qubit is equally coupled
to all the spins of the chain [$m=N$ in Eq.~\eqref{eq:interactionhamilt}].
This situation is opposite to the one considered in the previous subsection.
Interestingly, in this limit the model is almost solvable.
The system-bath Hamiltonian of Eq.~\eqref{eq:HamFree} reads:
\begin{equation}
  \mathcal{H}_{0}= -\omega_0 \ket{\downarrow} \bra{\downarrow}
  - J \sum_{j=1}^{N} \left[ \sigma_{j}^{x} \sigma_{j+1}^{x} +
    \left(\lambda +\frac{\epsilon}{J} \right) \sigma_{j}^{z} \right] \, .
\end{equation}
We first notice that
$\Ham_\downarrow (\lambda) = \Ham_\uparrow (\tilde{\lambda})
\equiv \Ham_E (\tilde{\lambda})$,
where $\tilde{\lambda}  = \lambda +\epsilon /J$. Therefore,
both the perturbed and the unperturbed Hamiltonians describe an Ising
model with a uniform transverse field, and can be diagonalized analytically
by means of a standard Jordan-Wigner-Fourier transformation, followed by
a Bogoliubov rotation.
Details on how to evaluate the LE of Eq.~\eqref{eq:LE-eff} for a spin-star
model can be found in Appendix~\ref{sec:Appendixeffectivedecay},
where we show that
\begin{equation}
  \left\langle G\right\vert e^{it\mathcal{H}_{\mathrm{eff}}} \ket{G} = 
  \prod_{k>0} \cos \left( 8 t \, \epsilon _{\mathrm{eff}}\Delta _{k} \right)\,,
  \label{eq:LE-final}
\end{equation}
with $\Delta_k = \sin (2 \pi k /N)$ and $\epsilon_{\mathrm{eff}}$
defined in Eq.~\eqref{eq:renormcc}.
In the limit of small $\epsilon _{\mathrm{eff}}$, while keeping $t$ finite,
we can approximate this as
\begin{equation}
  \left\langle G\right\vert e^{it\mathcal{H}_{\mathrm{eff}}} \ket{G} \simeq
  \prod_{k>0} e^{-\frac{1}{2}(8t \epsilon _{\mathrm{eff}}\Delta _{k})^{2}}
  = e^{-\frac{\Gamma }{2}(t\epsilon _{\mathrm{eff}})^{2}},
  \label{eq:G-ss}
\end{equation}
where we have defined
\begin{equation}
  \Gamma \equiv 64 \sum_{k>0}\Delta _{k}^{2} \, .
\end{equation}
%
\begin{figure}[!t]
\includegraphics[width=0.45 \textwidth]{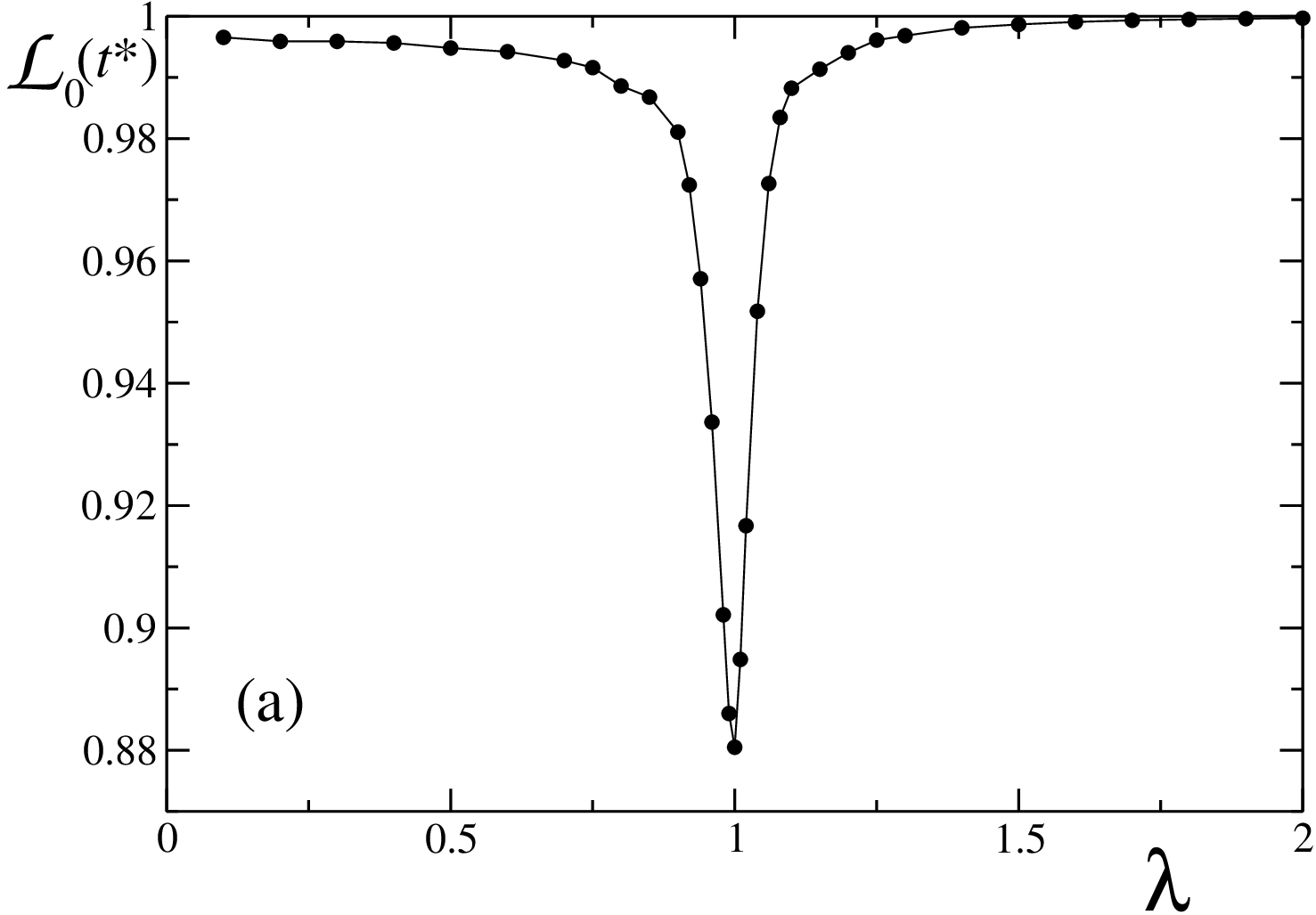}
\includegraphics[width=0.45 \textwidth]{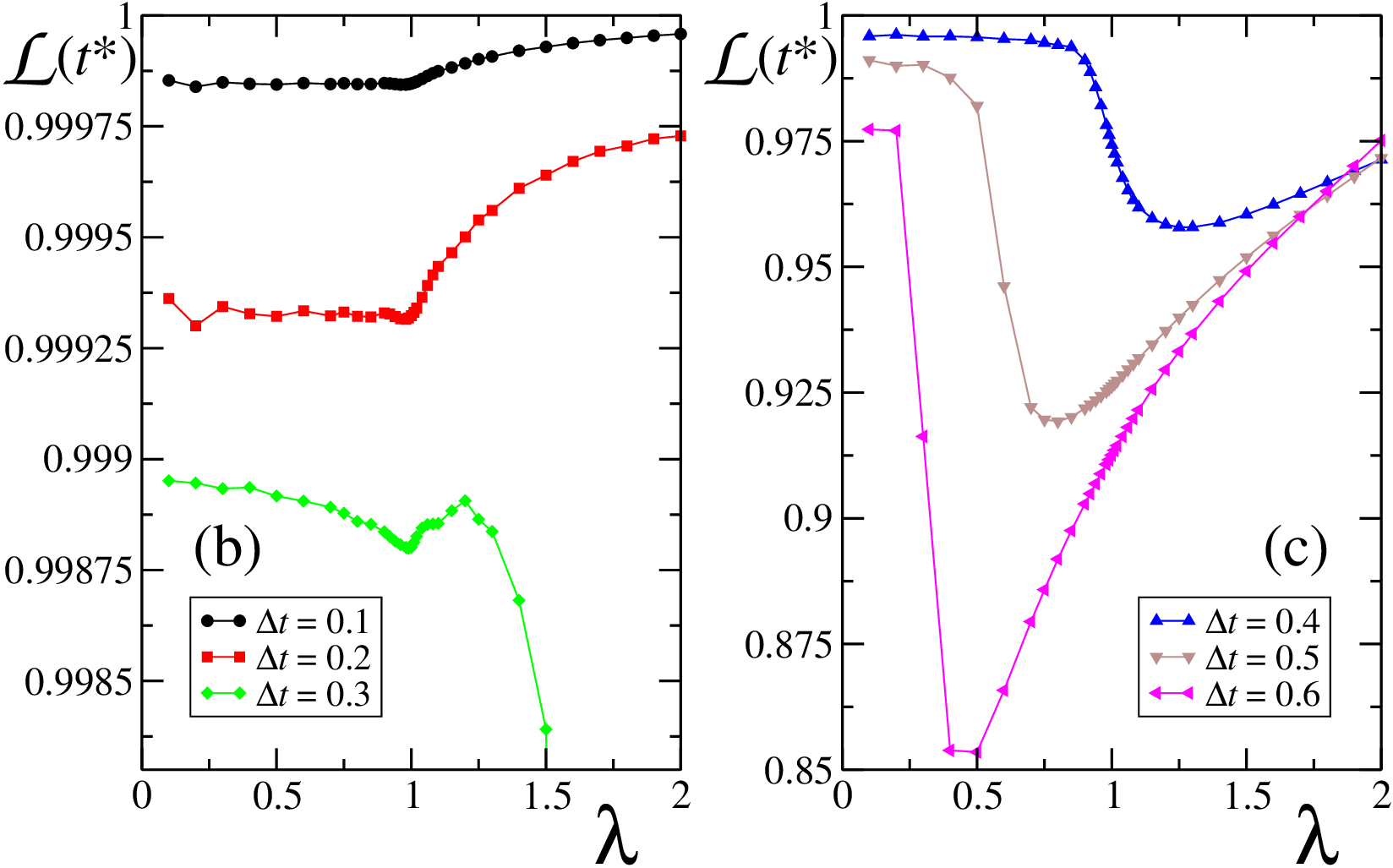}
\includegraphics[width=0.45 \textwidth]{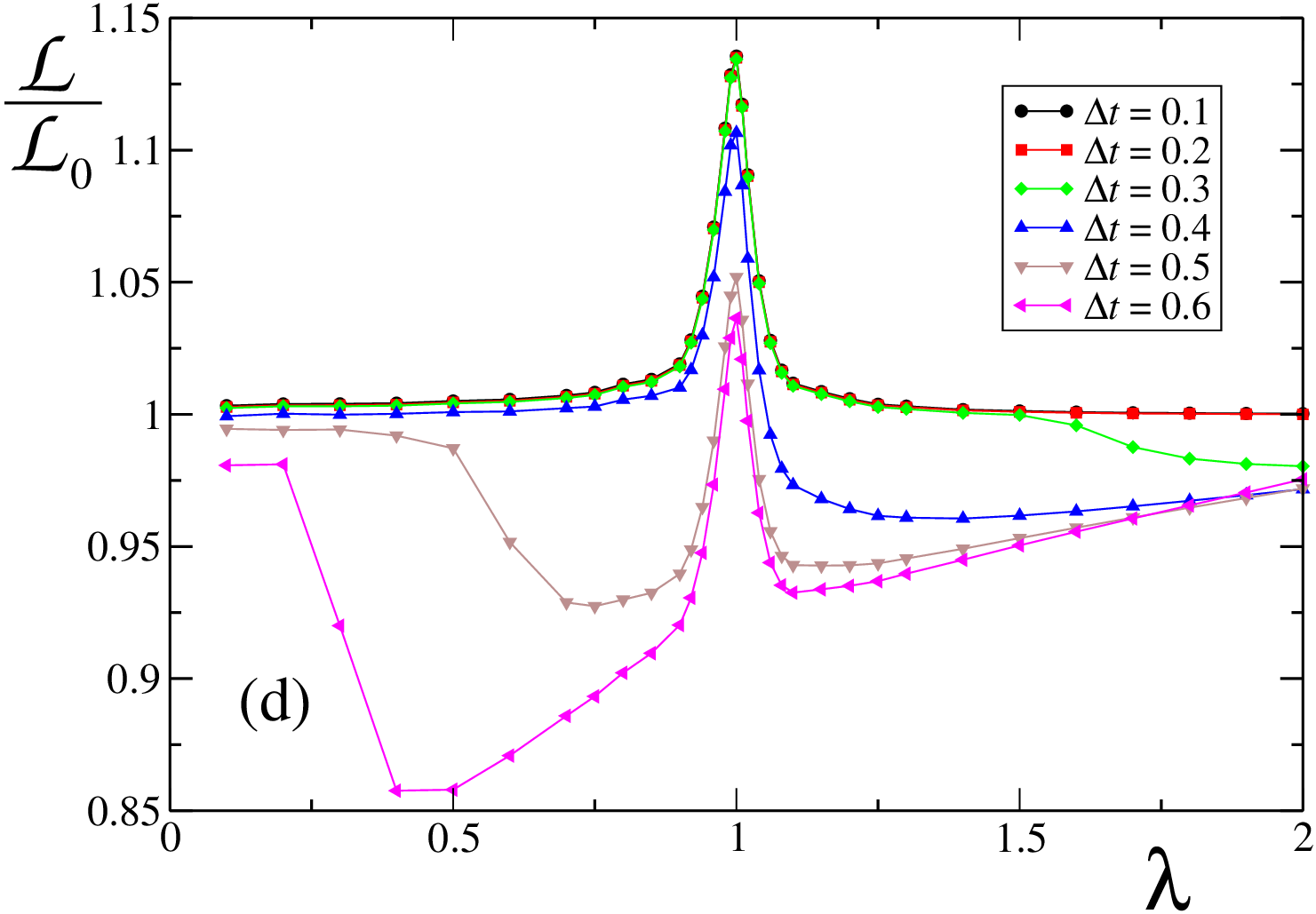}
\caption{(Color online) LE for the spin-star model (the parameters of the
  simulation are $J t^{\ast }=10$, $\protect\epsilon =0.01$ and $N=300$).
  (a): Dependence on $\protect\lambda $ of the LE without external
  control at fixed time. (b)-(c): LE in presence of pulsed control
  with frequency $\Delta t$. (d): Renormalized controlled LE.}
\label{fig:SpinStar}
\end{figure}
%
We notice that the dependence on $\lambda$ in Eq.~\eqref{eq:G-ss}
has entirely disappeared. This means that, to leading order in the
pulse interval $\Delta t$, dynamical decoupling is not sensitive to
criticality [Eq.~\eqref{eq:LE-eff} for the LE and
Eq.~\eqref{eq:LE-final} lead to $\mathcal{L} \approx e^{- \Gamma (t
\epsilon_{\rm eff})^2}$; we explicitly checked that, for small
$\Delta t$ and at short times, this formula exactly reproduces the
data obtained from numerical simulations, which are completely
insensitive to $\lambda$ in that regime]. This is consistent with
the data shown in Fig.~\ref{fig:SpinStar}. In panel (a) we see
the behavior of the LE in absence of control; we notice a slight dip
at $\lambda =1$. In panels (b)-(c) we study the LE for various
$\Delta t$; we observe strong similarities with Fig.~\ref{fig:frac},
in particular the weak dependence of the LE on $\lambda$ for very
small $\Delta t$. In panel (d) the rescaled LE again displays a cusp.

It is remarkable how similar the results are for $m=1$ (qubit coupled to a
single spin of the chain) and $m=N$ (spin-star model). The consistency of
these results and the analogies between these two opposite situations lead
us to conclude that general features of the decoherence of the qubit
under bang-bang control are largely independent of the number of chain spins
coupled to it, at least when the chain is close to criticality.

\section{Discussion and conclusions}

\label{sec:conclusion}

We have studied the efficacy of pulsed control of a qubit when it is
coupled to a spin bath. It is well known that, without control
pulses, the qubit decoheres particularly fast in the vicinity of the
critical point. The reason for this is that the evolution takes the
initial state $\ket{\Psi(0)}$, in the form of Eq.~\eqref{eq:InitialState},
into a superposition of the type $\ket{\uparrow}
\ket{\varphi_{0}(t)} + \ket{\downarrow} \ket{\varphi_{1}(t)}$
and the two bath states become rapidly orthogonal
near the critical point. The application of decoupling pulses to the
qubit removes the dependence of decoherence on the criticality of
the environment. On the other hand, we also found a regime (larger
interval $\Delta t$ between pulses) such that the control can
increase the effects of decoherence. Away from criticality the
perturbation is not as effective at orthogonalizing
$\ket{\varphi_{0}(t)}$ and $\ket{\varphi_{1}(t)}$, leading to a slow
decay of the echo and to relatively less effective control.
Therefore, we can conclude that in general decoupling is \emph{relatively}
more effective near the critical point, since there it results
in the largest enhancement of coherence.

From the quantum information processing perspective, there is
another positive message in these results: suppose we are trying to
preserve the coherence of a qubit in the presence of a spin bath.
Without decoupling we know that the spin decoheres particularly fast
in the vicinity of the critical point. Therefore not knowing whether
we are close to criticality when trying to operate a quantum computer
coupled to a spin bath, is a problem. But in light of the
results presented here, it follows that application of dynamical
decoupling pulses removes this concern: for sufficiently frequent
pulses, decoupling works independently of the value of the
system-bath coupling $\lambda$, so closeness to criticality does not
matter.

Our analytical and numerical calculations suggest that these results
seem to be largely independent of the details of the model of
qubit-environment coupling. Indeed, we have considered two extreme
situations (qubit coupled to a single spin of the chain and qubit
coupled to all spins in the chain), and obtained the same
qualitative behavior.

Finally, a comparison of different control strategies (Zeno effect,
decoupling pulses and strong continuous coupling)~\cite{FTPNTL2005} has
shown that, although these procedures are physically equivalent, there are
important practical differences among them. Future attention will be
directed towards the exploration of these similarities and differences in
the context of coupling of a qubit to a critical system.

\acknowledgments This work is partly supported by the European Community
through the Integrated Project EuroSQIP.
DAL was sponsored by the United States Department of Defense.
The views and conclusions contained in
this document are those of the authors and should not be interpreted
as representing the official policies, either expressly or implied, of
the U.S. Government.

\appendix

\section{}~\label{sec:Appendixechofree}

We explain here how to evaluate the LE for the Hamiltonian in
Eq.~\eqref{eq:HamFree}, and then extend some of these results to the case of
pulsed control, Eq.~\eqref{eq:HamGlobal}.
This technique can be easily generalized to the case
of an $XY$ spin bath, as has been done in Ref.~\cite{rossini07}.

By means of the Jordan-Wigner transformation
\begin{equation}
  \sigma _{j}^{+} = c_{j}^{\dagger }\exp \bigg( i \pi
  \sum_{k=1}^{j-1} c_{k}^{\dagger}c_{k} \bigg) \,, \quad
  \sigma _{j}^{z} = 2 c_{j}^{\dagger }c_{j}-1,
  \label{eq:JWT}
\end{equation}
we first map the Hamiltonians $\mathcal{H}_{\downarrow }$ and
$\mathcal{H}_{\uparrow }$ of the spin bath onto a free fermion model
that can be expressed in the form
\begin{equation}
  \mathcal{H}_{\uparrow /\downarrow }=\frac{1}{2}\mathbf{\Psi ^{\dagger}C\Psi},
\end{equation}
where $\mathbf{\Psi ^{\dagger }}=(c_{1}^{\dagger }\ldots c_{N}^{\dagger}
  \, c_{1}\ldots c_{N})$
($c_{i}$ being the corresponding spinless fermion operators) and
\begin{equation}
  \mathbf{C}=\sigma ^{z}\otimes \mathbf{A}+i\sigma ^{y}\otimes \mathbf{B}
\end{equation}
is a tridiagonal block matrix with
\begin{eqnarray}
  A_{j,k} &=&-J(\delta _{k,j+1}+\delta _{j,k+1})-2(\lambda +\epsilon
  _{j})\delta _{j,k} \\
  B_{j,k} &=&-J(\delta _{k,j+1}-\delta _{j,k+1})
\end{eqnarray}
such that $\epsilon _{j}=0$ for $\mathcal{H}_{\uparrow }$, while
$\epsilon_{j}=\epsilon \,\delta _{j,j_{m}}$ for $\mathcal{H}_{\downarrow }$.
The LE can then be evaluated exactly, by rewriting it in terms of
the determinant of a $2N\times 2N$ matrix~\cite{rossini07}:
\begin{equation}
  \mathcal{L}_{0}(t)=\left| \mathrm{det}\left( \mathbbm{1} -\mathbf{r}+\mathbf{r}\,
  e^{i\mathbf{C_{\downarrow }}t}\right) \right| \;,  \label{eq:LEcho}
\end{equation}
where $\mathbf{r}$ is a matrix whose elements $r_{i,j}=\langle \Psi_{i}^{\dagger}
\Psi _{j}\rangle $ are the two-point correlation functions of the spin chain,
evaluated in the ground state of the Hamiltonian $\mathcal{H}_{\uparrow }$.
Eq.~\eqref{eq:LEcho} can be obtained from the following trace formula~\cite{klich03}:
\begin{equation}
  \mathrm{Tr} \big( e^{\Gamma (A)}e^{\Gamma (B)} \big) =\mathrm{det}\left(
  \mathbbm{1} +e^{\mathbf{A}}e^{\mathbf{B}}\right) \,,  \label{eq:trace}
\end{equation}
where $\Gamma (A)=\sum_{i,j}\mathbf{A}_{ij}\,a_{i}^{\dagger }a_{j}$ and
$a_{i}^{\dagger },a_{i}$ are the creation and annihilation operators for a
fermion particle state $i$.

In the presence of pulsed control, in analogy with the free evolution case,
Eq.~\eqref{eq:LEchoFree}, we can rewrite the formula for the LE in
Eq.~\eqref{eq:LEchoPulse} in terms of the determinant of a $2N\times 2N$ matrix.
Indeed the trace formula~\eqref{eq:trace} is straightforwardly generalized to
products of more than two operators~\cite{klich03} by using the following identity:
\begin{eqnarray}
  &&\left\langle \psi _{0}\right\vert e^{-i\mathcal{H}_{1}t}
  e^{-i\mathcal{H}_{2}t}\ldots e^{-i\mathcal{H}_{n}t}\left\vert
  \psi _{0}\right\rangle  \notag \\
  &=&\mathrm{det}\left( \mathbbm{1}-\mathbf{r_{0}}+\mathbf{r_{0}}
  e^{-i\mathbf{C_{1}}t} e^{-i\mathbf{C_{2}}t}\ldots e^{-i\mathbf{C_{n}}t}\right) ,
\end{eqnarray}
where we supposed that $\mathcal{H}_{k}=\sum_{i,j}[\mathbf{C_{k}}]_{ij} \:
a_{i}^{\dagger }a_{j}$ and $\mathbf{r_{0}}=\Gamma (\mathcal{N})$ with
$\mathcal{N}$ occupation number operator
[i.e. $(\mathbf{r_{0}})_{ij}=\left\langle \psi_0 \right\vert
a_{i}^{\dagger }a_{j}\left\vert \psi_0 \right\rangle $].

\section{}~~\label{sec: AppendixLE}

We evaluate here the leading order expansion of the LE in
Eq.~\eqref{eq:LEchoPulse} in terms of the pulse interval $\Delta t$,
in the limit of short pulses. To simplify the notations, let us define
$A\equiv i\mathcal{H}_{\downarrow}$,
$B\equiv i\mathcal{H}_{\uparrow }$, and $\varepsilon \equiv \Delta t$.
We consider Eq.~\eqref{eq:LEchoPulse} at integer multiples of
a spin-flip cycle, i.e., $t=2 M \Delta t$:
\begin{equation}
  \mathcal{L}(t)=\Big\vert \mathrm{Tr} \left[\left\vert G\right\rangle \left\langle
    G\right\vert (e^{\varepsilon A} e^{\varepsilon B})^{M} (e^{-\varepsilon
      A}e^{-\varepsilon B})^{M}\right] \Big\vert^2
\end{equation}
Now recall the (approximate) Lie sum and product formulas
\begin{eqnarray}
  e^{\varepsilon A}e^{\varepsilon B} & = & e^{\varepsilon
    (A+B)}+O(\varepsilon^{2}), \\
  e^{\varepsilon A}e^{\varepsilon B}e^{-\varepsilon A}e^{-\varepsilon B} & = &
  e^{\varepsilon^{2} [A,B]}+O(\varepsilon^{3}).
\end{eqnarray}
Using this we have
\begin{eqnarray}
  & & (e^{\varepsilon A} e^{\varepsilon B})^{M} (e^{-\varepsilon A}
  e^{-\varepsilon B})^{M} =  \notag \\
  & = & (e^{\varepsilon A} e^{\varepsilon B})^{M-1} [e^{\varepsilon^{2} [A,B]}
    + O(\varepsilon^{3})] (e^{-\varepsilon A} e^{-\varepsilon B})^{M-1}  \notag
  \\
  & = & (e^{\varepsilon A} e^{\varepsilon B})^{M-2} [e^{\varepsilon^{2} [A,B]}
    e^{\varepsilon A} e^{\varepsilon B} + O(\varepsilon^{2})] (e^{-\varepsilon
    A} e^{-\varepsilon B})^{M-1}  \notag \\
  & & + O(\varepsilon^{3}) (e^{\varepsilon A} e^{\varepsilon B})^{M-1}
  (e^{-\varepsilon A} e^{-\varepsilon B})^{M-1}.
\end{eqnarray}
Keeping terms only to leading order $O(\varepsilon ^{2})$ we can neglect the
last line, yielding:
\begin{widetext}
  \begin{eqnarray}
  (e^{\varepsilon A} e^{\varepsilon B})^{M}(e^{-\varepsilon A} e^{-\varepsilon B})^{M}
  & = & (e^{\varepsilon A}e^{\varepsilon B})^{M-2} [e^{\varepsilon ^{2} [A,B]} e^{\varepsilon A} e^{\varepsilon B}
    e^{-\varepsilon A} e^{-\varepsilon B} + O(\varepsilon^{2})e^{-\varepsilon A}e^{-\varepsilon B}]
  (e^{-\varepsilon A} e^{-\varepsilon B})^{M-2} \nonumber\\
  & = & (e^{\varepsilon A} e^{\varepsilon B})^{M-2} [e^{2\varepsilon^{2} [A,B]} + O(\varepsilon ^{3})
    + O(\varepsilon ^{2}) (\mathbbm{1}-\varepsilon (A+B))] (e^{-\varepsilon A} e^{-\varepsilon B})^{M-2} \nonumber\\
  & = & (e^{\varepsilon A} e^{\varepsilon B})^{M-2}[e^{2 \varepsilon ^{2} [A,B]} + O(\varepsilon ^{2}) \mathbbm{1}]
  (e^{-\varepsilon A} e^{-\varepsilon B})^{M-2},
  \end{eqnarray}
\end{widetext}
where in the last line we again neglected $O(\varepsilon^{3})$ terms.
Continuing in this manner we have
\begin{equation}
  (e^{\varepsilon A} e^{\varepsilon B})^{M}(e^{-\varepsilon A} e^{-\varepsilon
    B})^{M} = e^{M\varepsilon^{2} [A,B]} + M O(\varepsilon^{2}) \mathbbm{1},
\end{equation}
which yields Eqs.~\eqref{eq:LE-eff}-\eqref{eq:H-eff}.

\section{}~~\label{sec:Appendixeffectivedecay}

Here we derive Eq.~\eqref{eq:LE-final}.
We first notice that, in the spin-star case, both
$\Ham_\uparrow (\lambda) \equiv \Ham_E (\lambda)$ and
$\Ham_\downarrow (\lambda) \equiv \Ham_E (\tilde{\lambda})$
can be written in momentum space, by using the
Jordan-Wigner transformation~\eqref{eq:JWT} followed by a
Fourier transform, in this way:
\begin{equation}
  \begin{array}{cl}
  \displaystyle \Ham_E (\lambda) = 2 J \sum_{k>0}
    & \hspace{-1.5mm} \displaystyle \big[
   \varepsilon_{k}(\lambda)(c_{k}^{\dag} c_{k} +c_{-k}^{\dag} c_{-k}) \\
   & \displaystyle - i\Delta_{k} (c_{k}^{\dag }c_{-k}^{\dag}-c_{-k}c_{k})
   \big] \end{array}
  \label{eq:Hferm}
\end{equation}
where $\varepsilon_{k}(\lambda) = \lambda - \cos ( 2 \pi k / N)$
and $\Delta_{k} = \sin (2 \pi k /N)$, and the sum over $k$ runs
from $1$ to $N/2$.

The ground state of the Hamiltonian in Eq.~\eqref{eq:Hferm} is
\begin{equation}
  \ket{G(\lambda)} =\bigotimes_{k>0} \left[
    \cos \Big( \frac{\theta_{k}}{2} \Big) \ket{00}_{k,-k}
    + i \sin \Big( \frac{\theta_{k}}{2} \Big) \ket{11}_{k,-k} \right] ,
\end{equation}
where $\theta_{k}=\arctan ( \Delta _{k} / \varepsilon_{k}(\lambda) )$,
and the kets refer to fermion occupation numbers in the two modes $k$ and $-k$.
Consider now the space
\begin{equation} \nonumber
  \mathbb{H}_{k} \otimes \mathbb{H}_{-k} = \mathrm{Sp}
  \{ \ket{00}_{k,-k},\ket{01}_{k,-k}, \ket{10}_{k,-k}, \ket{11}_{k,-k} \} \, .
\end{equation}
Since the subspaces $\mathrm{Sp}\{ \ket{00}_{k,-k}, \ket{11}_{k,-k} \}$
and $\mathrm{Sp}\{ \ket{01}_{k,-k}, \ket{10}_{k,-k} \}$ are not
coupled by $\mathcal{H}_E$, and since $\ket{G(\lambda )}$ lives in the
former two-dimensional subspace, we can rewrite the Hamiltonian
over the $\mathrm{Sp}\{\ket{00}_{k,-k},\ket{11}_{k,-k}\}$ subspace, up
to a constant, as
\begin{equation}
  \mathcal{H}_E (\lambda) = 2J \sum_{k>0} \left[ \varepsilon_{k}(\lambda) \Sigma_{k}^{z}
  + \Delta_{k} \Sigma_{k}^{y} \right]\equiv \sum_{k>0} \mathcal{H}_{E,k}(\lambda) \, ,
\end{equation}
where $\Sigma_{k}^{z}$ and $\Sigma_{k}^{x}$ generate an SU$(2)$ algebra and are
defined as
\begin{eqnarray}
  \Sigma_{k}^{x} & = &   c_{k}^{\dag} c_{-k}^{\dag} + c_{-k}        c_{k}  , \\
  \Sigma_{k}^{y} & = & -i(c_{k}^{\dag} c_{-k}^{\dag} - c_{-k}        c_{k}) , \\
  \Sigma_{k}^{z} & = &   c_{k}^{\dag} c_{k}         + c_{-k}^{\dag} c_{-k}-1 \, .
\end{eqnarray}
The problem of evaluating $\left\langle G\right\vert
e^{it\mathcal{H}_{\mathrm{eff}}}\left\vert G\right\rangle $ is now reduced
to computing the action of the $2\times 2$ matrix
$[\mathcal{H}_{\downarrow }, \mathcal{H}_{\uparrow }]$ over the subspace
$\mathrm{Sp}\{\ket{00}_{k,-k}, \ket{11}_{k,-k}\}$.
We can rewrite the ground state as
\begin{eqnarray}
  \ket{G(\lambda)} & = & \bigotimes_{k>0} \left[
    \cos \Big( \frac{\theta_{k}}{2} \Big) \ket{0}_{k}
    + i \sin \Big( \frac{\theta_{k}}{2} \Big) \ket{1}_{k} \right] \notag \\
  & \equiv & \otimes_{k>0} \ket{G_{k}(\lambda)} \, ,
\end{eqnarray}
where now $\ket{0}_{k}$ and $\ket{1}_{k}$ are the standard $\pm 1$
eigenstates of $\Sigma_{k}^{z}$.
Over this subspace, using the fact that
$\Ham_\downarrow (\lambda) = \Ham_\uparrow (\lambda) + \epsilon \Sigma^z$,
with $\Sigma^z = \sum_{j=1}^N \sigma^z_j$, we have that
\begin{eqnarray}
  \mathcal{H}_{\mathrm{eff}} & = & i \epsilon \frac{\Delta t}{2}
    [\Sigma^{z}, \mathcal{H}_{\uparrow}] =
    4 i \, \epsilon_{\mathrm{eff}} \sum_{k>0} \Delta_{k}[\Sigma_{k}^{z},\Sigma_{k}^{y}]  \notag\\
    & = & 8 \epsilon_{\mathrm{eff}} \sum_{k>0} \Delta_{k} \Sigma_{k}^{x}
    \equiv \sum_{k>0} \mathcal{H}_{\mathrm{eff}, k} \, .
\end{eqnarray}
Now,
\begin{equation}
  \Sigma_{k}^{x} \ket{G_{k}(\lambda)} =\bigg[ \cos \Big( \frac{\theta_{k}}{2} \Big)
    \ket{1}_{k} + i\sin \Big( \frac{\theta_{k}}{2} \Big) \ket{0}_{k} \bigg],
\end{equation}
so that
\begin{equation}
  \left\langle G_{k}\right\vert \Sigma_{k}^{x} \left\vert G_{k}\right\rangle = 0,
\end{equation}
and
\begin{eqnarray}
  \left\langle G_{k}\right\vert e^{i t \mathcal{H}_{\mathrm{eff}, k}}
  \left\vert G_{k}\right\rangle & = & \left\langle G_{k}\right\vert e^{8i t \,
    \epsilon_{\mathrm{eff}} \Delta _{k} \Sigma_{k}^{x}} \left\vert G_{k}\right\rangle
  \notag \\
  & = & \left\langle G_{k}\right\vert \cos (8 t \, \epsilon _{\mathrm{eff}} \Delta
  _{k}) \mathbbm{1}  \notag \\
  & & - i\sin (8 t \, \epsilon _{\mathrm{eff}} \Delta _{k}) \, \Sigma_{k}^{x} \left\vert
  G_{k}\right\rangle  \notag \\
  & = & \cos (8 t \, \epsilon_{\mathrm{eff}} \Delta_{k})
\end{eqnarray}
Therefore
\begin{eqnarray}
  \left\langle G\right\vert e^{it\mathcal{H}_{\mathrm{eff}}} \left\vert
  G\right\rangle & = & \Pi_{k>0} \left\langle G_{k}\right\vert e^{i t
    \mathcal{H}_{\mathrm{eff}, k}} \left\vert G_{k}\right\rangle  \notag \\
  & = & \Pi_{k>0} \cos (8 t \, \epsilon_{\mathrm{eff}} \Delta _{k}),
\label{eq:LE-finalapp}
\end{eqnarray}
which is Eq.~\eqref{eq:LE-final}.

\end{document}